\documentclass[a4paper,twocolumn,11pt,accepted=2023-02-13]{quantumarticle}
\pdfoutput=1

\usepackage{amsmath, amsfonts, amssymb}
\usepackage{hyperref}
\usepackage[utf8]{inputenc} \usepackage[T1]{fontenc}

\usepackage{graphicx}% Include figure files
\usepackage{dcolumn}% Align table columns on decimal point
\usepackage{bm}% bold math
\usepackage[colorlinks=true]{hyperref}% add hypertext capabilities
\usepackage{xspace}
\usepackage{physics}
\usepackage[export]{adjustbox}
\usepackage{mathtools}
\usepackage{amsmath,amssymb,amsfonts,stmaryrd}
\usepackage{environ}
\usepackage{amsthm}
\usepackage{relsize}
\usepackage{tikz-cd}

\usepackage{tikz}
\usetikzlibrary{quantikz}
\newcommand{\linedist}{0.55}

\renewcommand{\vec}[1]{\bm{#1}}

\newcommand{\oprod}{\bigotimes}
\newcommand{\etal}{\textit{et al}\xspace}

\begin{document}

	\title{Quantum circuits for solving local fermion-to-qubit mappings}% Force line breaks with \\
	
	\author{Jannes Nys}
    \email{jannes.nys@epfl.ch}
    \orcid{0000-0001-7491-3660}
    \affiliation{%
    \'{E}cole Polytechnique F\'{e}d\'{e}rale de Lausanne (EPFL), Institute of Physics, CH-1015 Lausanne, Switzerland
    }%
    \affiliation{%
    Center for Quantum Science and Engineering, \'{E}cole Polytechnique F\'{e}d\'{e}rale de Lausanne (EPFL), CH-1015 Lausanne, Switzerland
    }
    \author{Giuseppe Carleo}
    \orcid{0000-0002-8887-4356}
    \affiliation{%
    \'{E}cole Polytechnique F\'{e}d\'{e}rale de Lausanne (EPFL), Institute of Physics, CH-1015 Lausanne, Switzerland
    }%
    \affiliation{%
    Center for Quantum Science and Engineering, \'{E}cole Polytechnique F\'{e}d\'{e}rale de Lausanne (EPFL), CH-1015 Lausanne, Switzerland
    }
	
	
\begin{abstract}
Local Hamiltonians of fermionic systems on a lattice can be mapped onto local qubit Hamiltonians. Maintaining the locality of the operators comes at the expense of increasing the Hilbert space with auxiliary degrees of freedom. In order to retrieve the lower-dimensional physical Hilbert space that represents fermionic degrees of freedom, one must satisfy a set of constraints. 
In this work, we introduce quantum circuits that \emph{exactly} satisfy these stringent constraints. We demonstrate how maintaining locality allows one to carry out a Trotterized time-evolution with constant circuit depth per time step. Our construction is particularly advantageous to simulate the time evolution operator of fermionic systems in d>1 dimensions.  
We also discuss how these families of circuits can be used as variational quantum states, focusing on two approaches: a first one based on general constant-fermion-number gates, and a second one based on the Hamiltonian variational ansatz where the eigenstates are represented by parametrized time-evolution operators. 
We apply our methods to the problem of finding the ground state and time-evolved states of the $t$-$V$ model.
\end{abstract}
	
	\maketitle
	%\onecolumngrid
%	\tableofcontents
	
\section{Introduction}\label{sec:introduction}
Quantum simulations of fermionic systems on digital quantum devices can be challenging due to the anti-commutation relations of fermionic operators. More specifically, in order to carry out such simulations, one must map anti-commuting fermion operators onto Pauli  operators, which naturally obey commutation relations. 
The most commonly used mapping between fermionic and spin degrees of freedom is the Jordan-Wigner transformation (JWT), which follows as a natural consequence of the second quantization formalism of fermions~\cite{jordan1993paulische}. After choosing a fermion ordering in the second quantization formalism, the JWT maps each fermion operator $f_i^\dagger$ onto qubit operators $f_i^\dagger \to {\tfrac{1}{2}  (X_i-iY_i)} \left({\otimes_{j<i} Z_j}\right)$, where $(X_i, Y_i, Z_i)$ are Pauli matrices applied to qubit $i$. 
The operator chain $S_i = \otimes_{j<i} Z_j$ is commonly referred to as a Jordan-Wigner string, and is necessary to maintain the aforementioned fermionic anti-commutation relations. 

Typical physical Hamiltonians of fermions on a lattice (such as the Fermi-Hubbard model) consist of spatially local operators such as the hopping operators $f_i^\dagger f_j$, where $i$ and $j$ refer to fermionic modes on neighboring sites. Even though these operators are local on the fermionic side, under a JWT they can become non-local on the qubit operator side $f_i^\dagger f_j \to \tfrac{1}{4} (X-iY)_i (X+iY)_j \otimes_{i < k < j} Z_k$. The operator string $\otimes_{i < k < j} Z_k$ appears when the sites corresponding to modes $i$ and $j$ are not adjacent in the chosen JW ordering of the fermionic modes. For example, in the case of a 2D square lattice of size $L_x \times L_y$, the lattice sites are often assigned an order according to a horizontal snake-like pattern. In the latter case, vertical hopping terms will generate Jordan-Wigner strings with sizes up to $2L_x-2$~\cite{cade2020strategies}. The non-locality of the Jordan-Wigner strings become increasingly problematic as the dimensionality of the system increases~\cite{whitfield2016local, havlivcek2017operator}.

Fermion-to-qubit mappings can be avoided altogether in classical simulations using first quantization~\cite{hermann2022ab}, and in quantum simulation using analog simulators that natively contain fermionic degrees of freedom. Recent examples of such simulators are based on quantum dots~\cite{hensgens2017quantum, wang2022experimental} and cold atoms~\cite{brown2019bad}. However, there have recently been practical demonstrations of simulations of fermionic lattice systems with digital quantum processors. For example, Ref.~\cite{stanisic2022observing} successfully studied the time evolution of the 1d Fermi-Hubbard model, and Ref.~\cite{arute2020observation} studied ground-state properties of 1d and ladder systems. These practical implementations use traditional Jordan-Wigner encodings to simulate fermions using digital quantum devices. A number of recent theoretical works have focused on developing new methodologies to simulate 2d Jordan-Wigner-transformed fermionic systems on NISQ devices. Most prominently, techniques based on the use of Fermionic-SWAP gates (FSWAP) introduced in Ref.~\cite{kivlichan2018quantum}, and later elaborated upon in Ref.~\cite{cade2020strategies}, are promising for NISQ applications.  These approaches implement the non-local JW strings by swapping fermions to the edges of the lattice, where the hopping operators become local. Similarly, SWAP operators have been introduced to simulate fermionic systems in higher dimensions in second quantization with Tensor Networks~\cite{corboz2010simulation, orus2019tensor}. However, such methods require increasingly larger numbers of FSWAP operations when the system size or the dimensionality of the system increases, which is problematic for digital quantum simulation.

Fortunately, the JWT is not a unique mapping. This allows one to search, for example, for efficient encodings that require fewer qubits and/or reduce the operator weights~\cite{derby2021compact, jiang2020optimal, bravyi2002fermionic, steudtner2018fermion, setia2019superfast}, which can be beneficial on current intermediate-scale quantum devices that contain only a few tens of qubits~\cite{preskill2018quantum}. 
On the other hand, various works in the last $\sim$ 40 years have focused on generalizations to mappings in higher dimensions (>1d) with the aim to maintain \emph{locality} in the operators, and thereby reducing the size of the Jordan-Wigner strings. One of the first studies that derived higher-dimensional generalizations to the JWT dates back to the work of Wosiek~\cite{wosiek1981local}. Similar ideas were later explored independently by Ball~\cite{ball2005fermions} and Verstraete-Cirac~\cite{verstraete2005mapping} using methods that increase the fermionic Hilbert space with auxiliary fermions, while afterward restricting the reachable Hilbert space by defining an auxiliary Hamiltonian. Compared to Ref.~\cite{wosiek1981local}, Ball-Verstraete-Cirac transformations~\cite{ball2005fermions, verstraete2005mapping} suggested the more explicit introduction of auxiliary fermionic modes to counteract the Jordan-Wigner strings. These auxiliary modes effectively store the parity nearby the interaction terms, which is otherwise captured by the Jordan-Wigner string~\cite{whitfield2016local}. The auxiliary Majorana fermions are subject to local interaction terms that commute with the physical Hamiltonian, which is necessary in order to keep the eigenspectrum of the original problem identifiable in the spectrum of the transformed Hamiltonian. 

In recent years, we have witnessed a strong renewed interest in methods for simulating fermionic systems through a set of local qubit gates~\cite{po2021symmetric, setia2018bravyi, setia2019superfast, chen2018exact, chen2022equivalence, bochniak2020bosonization, li2021higher, po2021symmetric, derby2021compact}. Compared to earlier methods (such as Refs.~\cite{ball2005fermions, verstraete2005mapping}) where JWT was carried out explicitly, recent techniques take a different approach where one first defines bosonic operators from the fermionic ones, which can then be mapped directly onto qubit operators without the need to order the fermions. The equivalence of many of these bosonization procedures in 2D was proven recently by Chen~\etal\cite{chen2022equivalence}. Since most of these novel methods do not require one to order the fermionic modes, they can more easily be generalized to different systems and higher dimensions.

Despite the recent theoretical progress on fermion-to-qubit mappings, the practical application of these techniques to simulate 2d fermionic systems remains elusive, both in the context of classical computational methods (for which they were originally explored~\cite{wosiek1981local}), and as a basis for quantum algorithms. The main difficulty lies in the fact that the auxiliary degrees of freedom must satisfy stringent constraints in order to correctly represent fermionic degrees of freedom. Recently, however, we have studied local fermion-to-qubit mappings in >1d with classical variational techniques in Ref.~\cite{nys2022variational}. Hereby, we demonstrated that the quantum probability amplitudes function can be factorized into two parts: one part solving the Gauss constraints, and a variational part that only depends on the physical qubits (not on the auxiliary ones). However, this factorization is not unique. We investigated three solutions to satisfy the constraints, and showed that the complexity of the probability amplitudes function depends strongly on the chosen method to reduce the Hilbert space. Based on the insights of the most performant technique in Ref.~\cite{nys2022variational}, we design its quantum algorithm analog in this work. The result is a novel set of quantum circuits that can be used to study the recent local fermion-to-qubit mappings and bosonization procedures~\cite{ball2005fermions, verstraete2005mapping, po2021symmetric, setia2018bravyi, setia2019superfast, chen2018exact, chen2022equivalence, bochniak2020bosonization, li2021higher, po2021symmetric}, as well as its connected lattice gauge theories~\cite{chen2018exact}.

This paper is structured as follows. We discuss the general bosonization formalism in Section~\ref{sec:formalism} and discuss the emerging constraints. In Section~\ref{sec:vacuum_prep} we introduce a quantum circuit that satisfies all the constraints that follow from the bosonization. We provide an implementation of the time-evolution operator of the $t$-$V$ model using only local operators in Section~\ref{sec:time_evo}. We define two variational ansatzes: a first one based on general fixed-fermion-number rotations in Section~\ref{sec:variational_ansatz}, and a second one using the framework of Hamiltonian variational ansatzes based on the time-evolution operators in Section~\ref{sec:time_evo}. The performance of the variational ansatzes is demonstrated in Section~\ref{sec:vqe} where we determine the ground state of the $t$-$V$ model.
Finally, we compare our method to other state-of-the-art techniques in Section~\ref{sec:scaling} and discuss how to generalize it to other Hamiltonians, lattice topologies, internal degrees of freedom, and higher dimensions in Section~\ref{sec:generalization}.
	
\section{Formalism}\label{sec:formalism}
To introduce the formalism, we focus on spinless fermions on a square lattice. 
Consider a $L_x \times L_y$ 2D lattice with square cells, where all edges point either along the lattice basis vectors $\vec{x}$ or $\vec{y}$. 
This system consists of $N = \abs{\mathcal{V}}$ sites $\vec{r}\in \mathcal{V}$ connected with edges $\mathcal{E} = \{(\vec{r}, \vec{r}+\vec{x})| \vec{r} \in \mathcal{V}\} \cup \{(\vec{r}, \vec{r}+\vec{y})| \vec{r}\in \mathcal{V}\}$. Here, $\vec{r}=(r_x, r_y)$, where $r_x\in \{0,..., L_x-1\}$ and $r_x = L_x$ maps onto $r_x = 0$ due to periodicity (similar for $r_y$). 

In general, all bosonization procedures that maintain the locality of local fermion operators in >1d increase the number of qubits (either by introducing auxiliary fermions~\cite{verstraete2005mapping,ball2005fermions}, by generalizing the spin operators to generalized Euclidean spin operators~\cite{wosiek1981local}, or by considering qubits on the links~\cite{chen2018exact}). In the case of square lattices, each fermionic mode is associated with two qubits. We will refer to the physical (1) and auxiliary (2) qubit systems from hereon, and use their respective numbering to indicate operators that act on either system.
A second important consequence that follows from any bosonization procedure is that the increased Hilbert space must be reduced by fulfilling additional constraints. 
As pointed out by Chen \etal~\cite{chen2018exact}, a fermionic system on a 2D lattice can be mapped to a lattice gauge theory through bosonization procedures. The constraints can in this view be interpreted as the Gauss law of the lattice gauge theory~\cite{chen2018exact}
\begin{align}
    G_{\vec{r}} &\overset{c}{=} 1 \qquad \forall \vec{r} \in \mathcal{V}\label{eq:gausslaw}
\end{align}
The superscript $c$ indicates that this constraint is not an operator constraint, but must rather be satisfied by the allowed quantum states $\ket{\Psi}$, i.e.\ $G_{\vec{r}}\ket{\Psi} = \ket{\Psi}$, in order to maintain the fermionic properties.

We choose a bosonization procedure to determine $G_{\vec{r}}$ in terms of qubit operators. Here, we follow the bosonization recipe in Refs.~\cite{li2021higher, po2021symmetric}, but other auxiliary fermion procedures give similar representations (see e.g. Ref.~\cite{chen2018exact}) and are equivalent in 2D~\cite{chen2022equivalence}.
In particular, for square lattices, $G_{\vec{r}}$ in Eq.~\eqref{eq:gausslaw} takes the following bosonic (qubit) representation (see Ref.~\cite{nys2022variational, li2021higher, po2021symmetric} for a detailed derivation)
\begin{align}
    G_{\vec{r}} &= [Z^{(1)} Y^{(2)} ]_{\vec{r}} [X^{(2)} ]_{\vec{r}+\vec{x}} [Y^{(2)} ]_{\vec{r}+\vec{x}+\vec{y}} [Z^{(1)} X^{(2)})]_{\vec{r}+\vec{y}}\label{eq:gaussG} 
\end{align}
where the $G_{\vec{r}}$ commute for different $\vec{r}$. Here, the notation $O^{(s)}_{\vec{r}}$ is used to denote a Pauli operator $O$ applied to system $s$ at site $\vec{r}$.

We can separate the physical (1) and auxiliary (2) system and rewrite Eq.~\eqref{eq:gausslaw} as
\begin{align}
Y^{(2)}_{\vec{r}} X^{(2)}_{\vec{r}+\vec{x}} Y^{(2)}_{\vec{r}+\vec{x}+\vec{y}} X^{(2)}_{\vec{r}+\vec{y}} \overset{c}{=} Z^{(1)}_{\vec{r}}Z^{(1)}_{\vec{r}+\vec{y}}\label{eq:gaussrepr}
\end{align}
or in graphical form
\begin{align}
\begin{tikzpicture}
\node at (0,0) [rectangle,draw, style={draw, inner sep=0pt, minimum size=0.7cm}, rounded corners=0.1cm, fill={rgb:black,1;white,2}] (r) {$Y^{(2)}$};
\node at (1,0) [rectangle,draw, style={draw, inner sep=0pt, minimum size=0.7cm}, rounded corners=0.1cm, fill={rgb:black,1;white,2}] (rx) {$X^{(2)}$};
\node at (1,1) [rectangle,draw, style={draw, inner sep=0pt, minimum size=0.7cm}, rounded corners=0.1cm, fill={rgb:black,1;white,2}] (rxy) {$Y^{(2)}$};
\node at (0,1) [rectangle,draw, style={draw, inner sep=0pt, minimum size=0.7cm}, rounded corners=0.1cm, fill={rgb:black,1;white,2}] (ry) {$X^{(2)}$};
\draw (r) -- (rx) -- (rxy) -- (ry) -- (r);
\node at (1.75,0.5) [] (e) {$\overset{c}{=}$};
\node at (3.5,1) [circle,fill,inner sep=1.5pt] (rxy1) {};
\node at (3.5,0) [circle,fill,inner sep=1.5pt] (rx1) {};
\node at (2.5,0) [rectangle,draw, style={draw, inner sep=0pt, minimum size=0.7cm}, rounded corners=0.1cm, fill={rgb:black,1;white,2}] (r1) {$Z^{(1)}$};
\node at (2.5,1) [rectangle,draw, style={draw, inner sep=0pt, minimum size=0.7cm}, rounded corners=0.1cm, fill={rgb:black,1;white,2}] (ry1) {$Z^{(1)}$};
\draw (r1) -- (rx1) -- (rxy1) -- (ry1) -- (r1);
\end{tikzpicture} \nonumber
\end{align}
The operators 
\begin{align}
    C_{\vec{r}} =Y^{(2)}_{\vec{r}} X^{(2)}_{\vec{r}+\vec{x}} Y^{(2)}_{\vec{r}+\vec{x}+\vec{y}} X^{(2)}_{\vec{r}+\vec{y}} \label{eq:gaussCdef}
\end{align}
on the left-hand side are also the Hamiltonian operators in Wen's plaquette model~\cite{wen2003quantum, bonilla2020xzzx}. The fact that the right-hand side depends on the physical system, however, makes this a dynamical version of the plaquette model. As pointed out in Ref.~\cite{chen2018exact}, these constraints resemble a Chern-Simons Gauss law, which captures flux attachment. The Chern-Simons gauge theories that attach magnetic flux and electric charges are typically challenging.
For simplicity, we will always refer to Eq.~\eqref{eq:gausslaw} as the Gauss constraints from hereon. 

By imposing periodic boundary conditions in the fermionic system, we obtain additional constraints introduced by non-contractable Wilson loops that must be satisfied by valid quantum states~\footnote{We focus on the case where $L_x$ and $L_y$ are either both odd or both even.} (see again Refs.~\cite{nys2022variational, li2021higher, po2021symmetric} for a detailed derivation)
\begin{align}
     \oprod_{m=0}^{L_x-1} Z^{(1)}_{\vec{r}+m\vec{x}} Z^{(2)}_{\vec{r}+m\vec{x}} &\overset{c}{=} -1 \label{eq:pbc_constraint_x} \\      
     \oprod_{m=0}^{L_y-1} Z^{(2)}_{\vec{r}+m\vec{y}} &\overset{c}{=} -1 \label{eq:pbc_constraint_y}
\end{align}
We will refer to these as the periodicity constraints.

\section{Satisfying the constraints}\label{sec:vacuum_prep}

We demonstrate how to construct variational quantum circuits that conserve the number of fermions, obey the parity constraints in Eqs.~\eqref{eq:pbc_constraint_x} and \eqref{eq:pbc_constraint_y}, and fulfill the Gauss constraint in Eq.~\eqref{eq:gausslaw} everywhere in its parameter domain. 

\begin{figure}[bt]
\centering
\begin{tikzpicture}[scale=0.7]
\def\ngrid{4};
\def\ymax{\ngrid+1};
\def\xmax{\ngrid+1};
\def\gatewidth{0.4};
  \foreach \x in {1,...,\ngrid}
      \draw (\x , 0)--(\x , \ymax) ;
  \foreach \y in {1,...,\ngrid}
      \draw (0 , \y)--(\xmax , \y) ;
      
  \foreach \x in {1,...,\ngrid}
  	\foreach \y in {1, ..., \ngrid}
  		\node at (\x, \y) [circle,fill,inner sep=1.5pt]{};
  \foreach \i in {0, 1, 2, 3} {
    	\draw[fill=white, rounded corners=0.1cm, fill={rgb:black,1;white,2}] (\i+1-\gatewidth, \i+1-\gatewidth) rectangle +(2*\gatewidth, 2*\gatewidth) node[pos=.5] {$X$};
   }
\end{tikzpicture}
\caption{Example of how a periodicity circuit $V_P$ applies $X^{(2)}$ gates to the auxiliary system for a $4 \times 4$ periodic square lattice, in order to satisfy Eq.~\eqref{eq:pbc_constraint_x} and~\eqref{eq:pbc_constraint_y}.
\label{fig:periodicity_circuit}}
\end{figure}
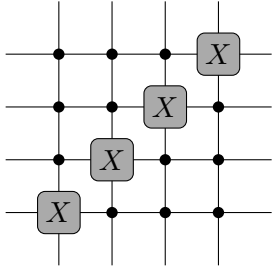

\begin{figure}[tb]
\centering
\includegraphics[width=0.45\textwidth, trim={0cm 24cm 45cm 0cm},clip]{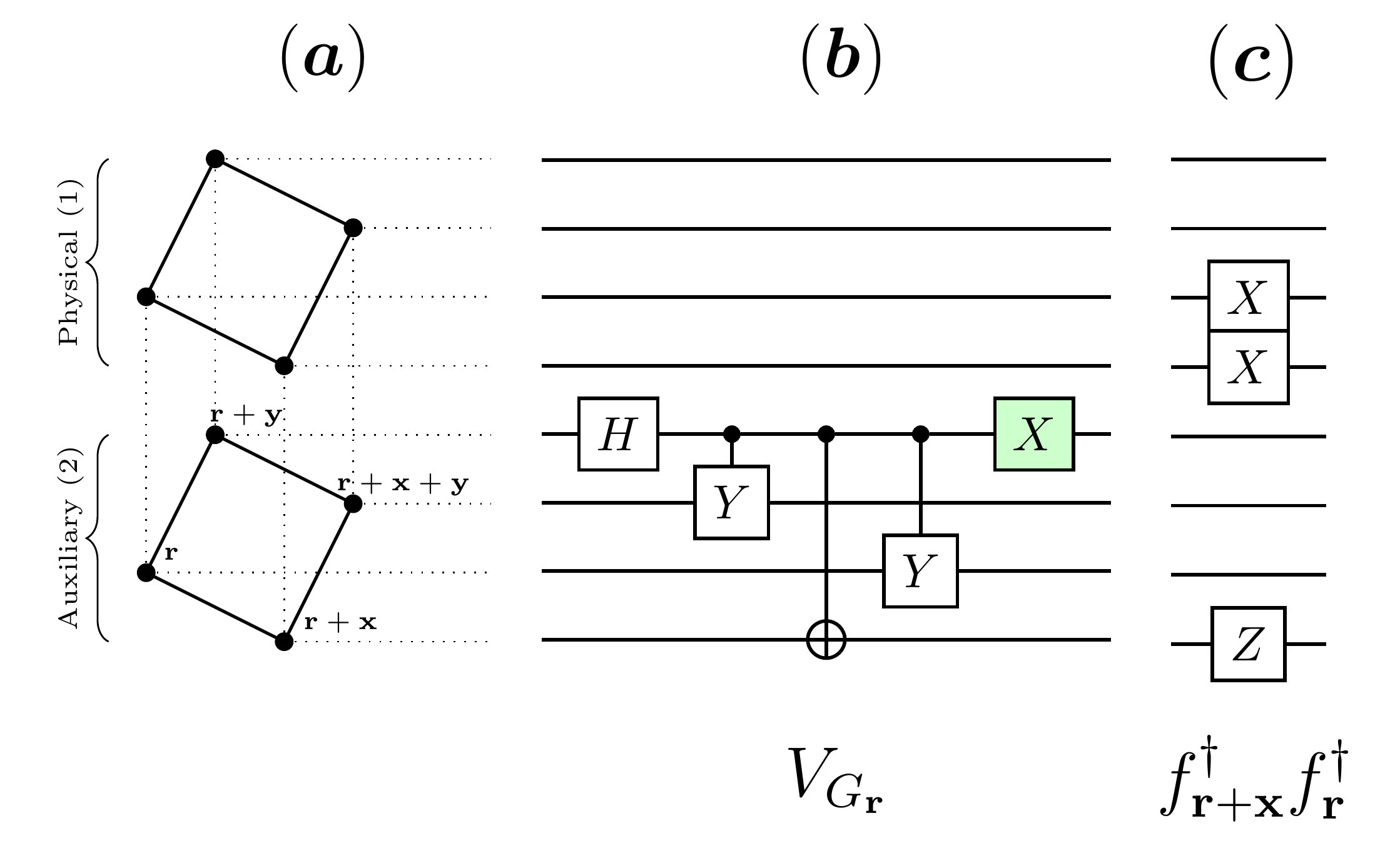}
\caption{Illustration of the building blocks of the initial-state quantum circuit for a single plaquette, excluding the periodic circuit $V_P$. From left to right. (a) Identification of the qubit lines with the physical and auxiliary qubit grid. (b) Vacuum preparation circuit $V_{G_{\vec{r}}}$, where the green $X$ gate is only applied if the periodic circuit $V_P$ would insert an $X$ gate. (c) Inserting a pair of fermions in positions $\vec{r}$ and $\vec{r}+\vec{x}$.  \label{fig:plaqcircuit_initial}}
\end{figure}

\subsection{Vacuum-state preparation}
The vacuum state $\ket{\Omega_G}$ is prepared such that it fulfills the Gauss~\eqref{eq:gausslaw} and periodicity constraints \eqref{eq:pbc_constraint_x}--\eqref{eq:pbc_constraint_y}. We will later define operators that only transform states without leaving the space defined by these constraints.

We start from the $\ket{0}^{\otimes 2N}$ state. At this point the global periodicity constraints \eqref{eq:pbc_constraint_x}--\eqref{eq:pbc_constraint_y} can be easily fulfilled with an operator $V_P$ working on the auxiliary system only. In this case, the constraints reduce to operators on the auxiliary system: \begin{align}
     \oprod_{m=0}^{L_x-1} Z^{(2)}_{\vec{r}+m\vec{x}} \overset{c}{=} -1,  \quad \oprod_{m=0}^{L_y-1} Z^{(2)}_{\vec{r}+m\vec{y}} \overset{c}{=} -1 
\end{align}
Hence, for example for $L_x = L_y$ even lattices, the constraints are satisfied by applying $X$ gates to the auxiliary qubits on the grid diagonal (see Figure~\ref{fig:periodicity_circuit}).

Next, we focus on satisfying the Gauss-law constraint~\eqref{eq:gausslaw}. 
We introduce a circuit that \emph{exactly} satisfies the challenging constraint as well.
An overview of the circuit operations described below is shown in Figs.~\ref{fig:plaqcircuit_initial} and \ref{fig:plaqcircuit_variational} for a single plaquette attached to $\vec{r}$. We identify each plaquette (containing both the physical and auxiliary qubits plaquette) with the bottom-left site $\vec{r}$. The physical system of the periodic state $V_P \ket{0}$ remains in the $\ket{0}$ state, and therefore, also the operator $G_{\vec{r}}$ reduces to a non-dynamical constraint on the auxiliary system only: $G_{\vec{r}} V_P \ket{0} = C_{\vec{r}} V_P \ket{0}$, where the plaquette operator $C_{\vec{r}}$ is defined in Eq.~\eqref{eq:gaussCdef}. Hence, we introduce a circuit that exactly produces an eigenstate of Wen's plaquette model~\cite{wen2003quantum}.

A state satisfying $C_{\vec{r}}V_P\ket{0} = V_P\ket{0}$ can be prepared by a set of conditional gates in Figure~\ref{fig:plaqcircuit_initial}b.  The corresponding circuit is denoted by $V_{G_{\vec{r}}}$. In order not to apply the Hadamard gate to auxiliary qubits in the $\ket{1}$ state, we apply the $X$ gate on the conditioning qubit $q_{\vec{r}+\vec{y}}$ after applying $V_{G_{\vec{r}}}$. Since each application of $V_{G_{\vec{r}}}$ flips the state of pairs of qubit along each grid line, the periodicity constraints remain automatically satisfied. The operator $V_{G_{\vec{r}}}$ is applied to all plaquettes sequentially, resulting in the circuit $V_G = \oprod_{\vec{r}\in \mathcal{V}} V_{G_{\vec{r}}}$. Since each $V_{G_{\vec{r}}}$ can be implemented using three conditional two-qubit gates (and a Hadamard gate), the cost of implementing $V_G$ is $3(L_x - 1)(L_y-1)$ two-qubit gates, i.e.\ $\order{N}$. We denote the resulting vacuum state as $\ket{\Omega_G} = V_{G} V_P \ket{0}$. Up to this point, the current vacuum state does not depend on the chosen Hamiltonian, and contains no variational parameters. 
% In Appendix~\ref{sec:compilation_gauss} we study the possibility of compiling the state $\ket{\Omega_G}$ with a variational circuit. 
% The simplest imaginable approach to this problem is to determine $\ket{\Omega_G} = V_G(\theta) V_P \ket{0}$ with a variational circuit $V_G(\theta)$. The parameters $\theta$ are obtained by minimizing the augmented Hamiltonian which consist of the system Hamiltonian, with an additional auxiliary term $H_{aux} = -K\sum_{
% \vec{r}} G_{\vec{r}}$ with $K \to +\infty$. This state would in most cases only approximately satisfy the Gauss law constraint in Eq.~\eqref{eq:gausslaw}. Indeed, this approach does not work well in practice, and therefore,

\subsection{Creating fermions} 
We now insert pairs of fermions into the vacuum $\ket{\Omega_G}$, while maintaining the constraints above. We respect the periodicity and Gauss constraints by inserting pairs of fermions through the bosonized version of the corresponding fermionic operators $f^\dagger_i f^\dagger_j$ (we remind the reader that there is no need to define an order in the fermionic modes, except for the direction of the lattice edges). For example, we find the following qubit operator representations for pairwise creation operators on neighboring sites 
\begin{align}
f^\dagger_{\vec{r}+\vec{x}} f^\dagger_{\vec{r}} &\to  Q_{\vec{r}}^{+(1)} Q_{\vec{r}+\vec{x}}^{+(1)} Z^{(2)}_{\vec{r}+\vec{x}} \label{eq:creation_pair_x}
\end{align}
where $Q^{\pm (1)}_{\vec{r}} = (X^{(1)}_{\vec{r}} \mp i Y^{(1)}_{\vec{r}})/2$.  When applied to the vacuum $\ket{\Omega_G}$, the pair creation operator in Eq.~\eqref{eq:creation_pair_x} reduces to $X^{(1)}_{\vec{r}} X^{(1)}_{\vec{r}+\vec{x}} Z^{(2)}_{\vec{r}+\vec{x}} \ket{\Omega_G}$, and can be implemented with the set of single-qubit gates in Fig.~\ref{fig:plaqcircuit_initial}c. Most notable is that we require a gate operation on the auxiliary system, which stores the parity of the system. One can similarly insert pairs of fermions on sites connected by vertical edges:
\begin{align}
f_{\vec{r}+\vec{y}}^\dagger f_{\vec{r}}^\dagger &\to  Q_{\vec{r}}^{+(1)} Q_{\vec{r}+\vec{y}}^{+(1)} Y^{(2)}_{\vec{r}}X^{(2)}_{\vec{r}+\vec{y}}  \label{eq:creation_pair_y}
\end{align}

\section{Time-evolution operator}\label{sec:time_evo}
We will focus on the $t$-$V$ model (often referred to as the spinless Fermi-Hubbard model). By taking a system without internal spin degrees of freedom, a system of moderate size can be computed exactly on classical hardware. On the fermionic side, the Hamiltonian reads,
\begin{align}
\begin{split}
    H = &-t \sum_{(\vec{r}, \vec{r}') \in \mathcal{E}} \left[f^\dagger_{\vec{r}} f_{\vec{r}'} + f^\dagger_{\vec{r}'} f_{\vec{r}} - \frac{V}{t} n_{\vec{r}} n_{\vec{r}'} \right] \label{eq:hamiltonian_fermion}
\end{split}
\end{align}
where we have introduced the usual number operator $n_{\vec{r}} = f^\dagger_{\vec{r}} f_{\vec{r}}$. Bosonizing the Hamiltonian in Eq.~\eqref{eq:hamiltonian_fermion} with the procedure in Ref.~\cite{li2021higher} results in:
\begin{align}
\begin{split}
    H =& -\frac{t}{2} \sum_{\vec{r}\in \mathcal{V}}   ( X^{(1)}_{\vec{r}}X^{(1)}_{\vec{r}+\vec{x}} + Y^{(1)}_{\vec{r}}Y^{(1)}_{\vec{r}+\vec{x}})Z^{(2)}_{\vec{r}+\vec{x}} \\ 
    &-\frac{t}{2} \sum_{\vec{r}\in \mathcal{V}} 
    (-X^{(1)}_{\vec{r}}Y^{(1)}_{\vec{r}+\vec{y}} + Y^{(1)}_{\vec{r}}X^{(1)}_{\vec{r}+\vec{y}}
    )Y^{(2)}_{\vec{r}} X^{(2)}_{\vec{r}+\vec{y}}  \\
    &+ \frac{V}{4} \sum_{(\vec{r},\vec{r}') \in \mathcal{E}} (1-Z^{(1)}_{\vec{r}})(1-Z^{(1)}_{\vec{r}'}) \label{eq:spinless_ham_boson}
\end{split}
\end{align}
As one can clearly observe, this Hamiltonian now only contains local operators that operate on qubits corresponding to lattice sites that are connected by an edge. 

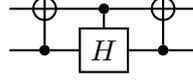
\begin{figure}[bt]
\begin{quantikz}[column sep=0.3cm, row sep={\linedist cm,between origins}]
 & \targ{} & \ctrl{1} & \targ{}  & \qw \\
 & \ctrl{-1} & \gate{H} & \ctrl{-1} & \qw
\end{quantikz}
\caption{Operator $W$ for mapping $XX + YY$ onto $ZI - IZ$, defined in Ref.~\cite{cade2020strategies}\label{fig:hopping_map}}
\end{figure}

\begin{figure}[tb]
\begin{quantikz}[column sep=0.3cm, row sep={0.7 cm,between origins}]
& \qw & \qw & \targ{} & \gate{R_Z(\theta)} & \targ{} & \qw & \qw & \qw \\
& \gate[]{R_X(\text{-}\tfrac{\pi}{2})} & \targ{} & \ctrl{-1} & \qw & \ctrl{-1} & \targ{} & \gate[]{R_X(\tfrac{\pi}{2})} &\qw \\
& \gate[]{R_Y(\tfrac{\pi}{2})} & \ctrl{-1} & \qw & \qw & \qw & \ctrl{-1} & \gate[]{R_Y(\text{-}\tfrac{\pi}{2})} & \qw
\end{quantikz}
\caption{Circuit to implement $e^{-i \theta Z Y X}$.
\label{fig:hopping_y_evol}}
\end{figure}
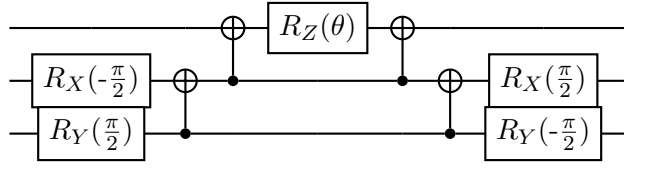

In order to study the time evolution of a fermionic system under the Hamiltonian of the $t$-$V$ model, we introduce the corresponding time-evolution operator $e^{-iHt}$. 
The Hamiltonian in Eq.~\eqref{eq:hamiltonian_fermion} contains three terms: horizontal and vertical hopping terms, and Coulomb interaction terms: $H = H_{\vec{x}} + H_{\vec{y}} + H_{C}$. Using Trotterization, we can approximate the time evolution with the  operators $e^{-iH \delta t} \approx e^{-iH_C \delta t}e^{-iH_{\vec{x}} \delta t}e^{-iH_{\vec{y}} \delta t}$ for sufficiently small $\delta t$.
Hence, it is sufficient that we define the time-evolution operators for the three Hamiltonian terms individually. 

We first focus on the horizontal hopping term. We map the physical operators $X^{(1)}_{\vec{r}} X^{(1)}_{\vec{r}+\vec{x}} + Y^{(1)}_{\vec{r}} Y^{(1)}_{\vec{r}+\vec{x}}$ to $Z^{(1)}_{\vec{r}}-Z^{(1)}_{\vec{r}+\vec{x}}$~\cite{cade2020strategies}.
The latter can be accomplished with the transformation $W$ in Fig.~\ref{fig:hopping_map}, i.e.\ $XX + YY = W(ZI -IZ)W^\dagger$. We will denote $W$ as $W_{\vec{x}}$ when it is applied to a horizontal link. Within this basis, the time evolution under the horizontal hopping can be implemented by a sequential application of the operators $e^{i \theta Z^{(1)}_{\vec{r}} Z^{(2)}_{\vec{r}+\vec{x}}}$ and $e^{-i \theta Z^{(1)}_{\vec{r}+\vec{x}} Z^{(2)}_{\vec{r}+\vec{x}}}$, followed by uncomputing the operator $W_{\vec{x}}$. 

Similarly, we can use the mapping $W_{\vec{y}} = W S_{\vec{r}}$ to map the physical part of the vertical hopping operator onto $Z^{(1)}_{\vec{r}} - Z^{(1)}_{\vec{r}+\vec{y}}$. Hereby, the $S$ gate transforms $-XY+YX$ into $XX+YY$ first. Notice that the auxiliary system operator $Y^{(2)}_{\vec{r}} X^{(2)}_{\vec{r}+\vec{y}}$ can be diagonalized using a map of single-qubit gates $R_X(-\tfrac{\pi}{2})\otimes R_Y(\tfrac{\pi}{2})$. The time-evolution then requires evolving the system according to $e^{i\theta Z^{(1)}_{\vec{r}} Z^{(2)}_{\vec{r}}Z^{(2)}_{\vec{r}+\vec{y}} }$ and $e^{-i\theta Z^{(1)}_{\vec{r}+\vec{y}} Z^{(2)}_{\vec{r}}Z^{(2)}_{\vec{r}+\vec{y}} }$, similarly as for the horizontal case. The circuit implementing this rotation is depicted in Fig.~\ref{fig:hopping_y_evol}.
The second evolution operator does not require additional CNOTs and $R_X \otimes R_Y$ gates in Fig.~\ref{fig:hopping_y_evol}, since the same basis and parity from the auxiliary system can be used. Hence, the second factor only requires one to insert a $R_Z(-\theta)$ gate surrounded by two CNOT gates that target the $q^{(1)}_{\vec{r}+\vec{y}}$ physical qubit in Fig.~\ref{fig:hopping_y_evol}. The three-qubit rotation operator is implemented with the circuit in Fig.~\ref{fig:hopping_y_evol}.

Lastly, the interaction terms $\tfrac{1}{4}(1-Z^{(1)}_{\vec{r}})(1-Z^{(1)}_{\vec{r}'})$ operate only on the physical qubits. Therefore, their circuits remain the same as with the standard JWT, which is given by a controlled phase-shift gate $CR_{\phi}(\theta)$~\cite{arute2020observation}.

\begin{figure}[tbh]
    \centering
    \includegraphics[width=0.45\textwidth]{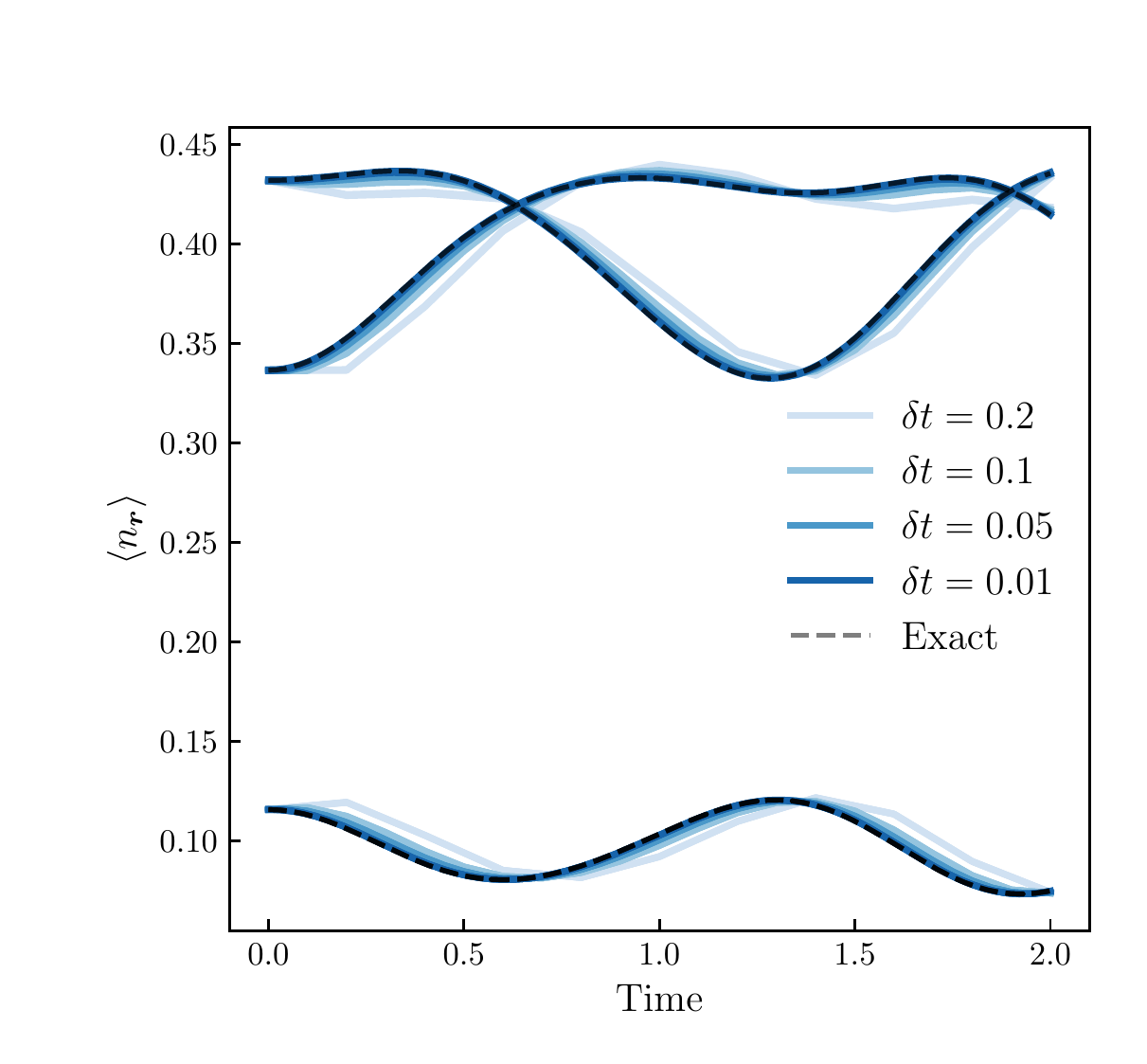}
    \caption{Expectation value of the occupation $\expval{n_{\vec{r}}}$ of site $\vec{r}$ as a function of time, using our Trotterized circuit with time steps $\delta t$. Exact dynamics are obtained with NetKet~\cite{vicentini2021netket}}
    \label{fig:time_evo}
\end{figure}

We demonstrate the correctness of the circuit in the context of the real-time evolution of a state. As initial state, we take $N_f=2$ subject to the free-fermions model ($V/t = 0$) with an additional potential $-K (n_{\vec{0}} + n_{\vec{y}})$ with $K=1$. The system is quenched and evolves under the $t$-$V$ model in Eq.~\eqref{eq:hamiltonian_fermion} with $V/t = 3$ (i.e.\ $K=0$), with time step $\delta t$.  The results are shown in Fig.~\ref{fig:time_evo}.

\section{Variational state preparation}\label{sec:variational_ansatz}
In Section~\ref{sec:vacuum_prep}, we introduced quantum circuits to satisfy ($i$) the periodicity constraint, ($ii$) Gauss constraint, and ($iii$) the fermion-number constraint. In this section, we will focus on introducing variational circuits that maintain these constraints but offer the flexibility to represent physical quantum states.

\begin{figure*}[tb]
\centering
\includegraphics[width=0.9\textwidth, trim={0cm 24cm 20cm 0cm},clip]{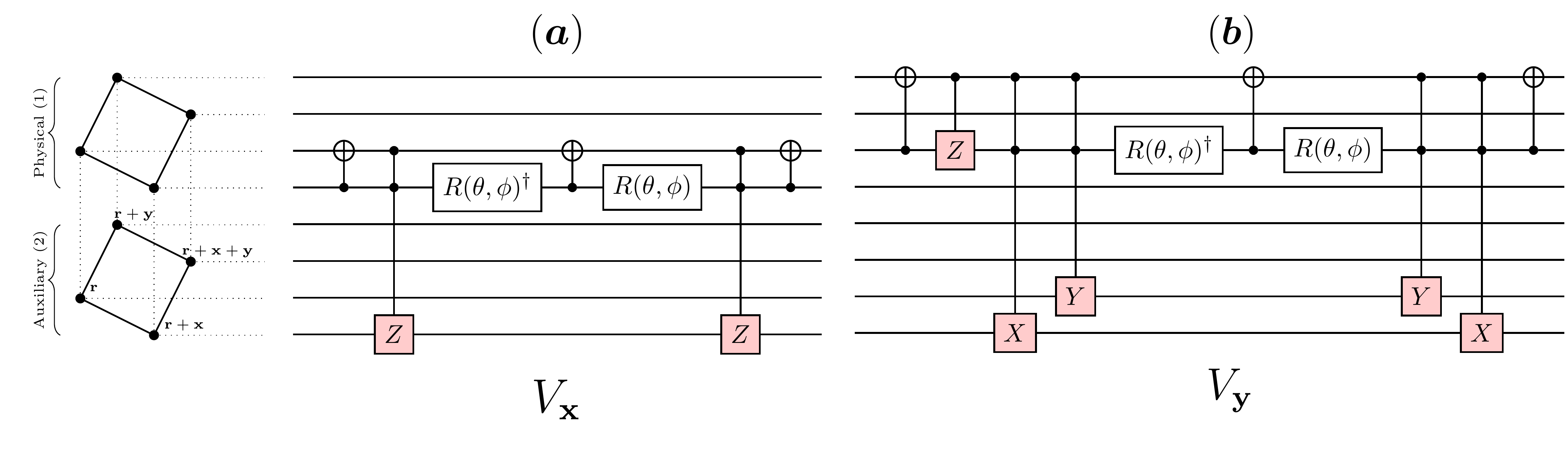}
\caption{Illustration of the components of the variational circuit for a single plaquette, as in Fig.~\ref{fig:plaqcircuit_initial}. (a) Hopping operators along the horizontal $\vec{x}$ axis, where $R(\theta, \phi) = R_Z(\phi+\pi) R_Y(\theta+\tfrac{\pi}{2})$. (b) Hopping operators along the vertical $\vec{y}$ axis. The red gates indicate the additional gates compared to the approach in Ref.~\cite{barkoutsos2018quantum, gard2020efficient}.\label{fig:plaqcircuit_variational}}
\end{figure*}

\subsection{Fixed-particle-number rotations}\label{sec:fixed_Nf_rot}

We provide two circuits that implement the variational part, one using the framework of Ref.~\cite{gard2020efficient} based on fixed particle-number rotations in the current section, and a second one based on the Trotterized time-evolution in Section~\ref{sec:time_evo}, which is useful from the perspective of the Hamiltonian variational (HV) ansatz~\cite{wecker2015progress, cade2020strategies}.

We fix the number of fermions to be $N_f$, which is enforced through the constraint
\begin{align}
    \sum_{\vec{r}\in \mathcal{V}}\frac{1-Z^{(1)}_{\vec{r}}}{2} \overset{c}{=} N_f \label{eq:Nf_constraint}
\end{align}
Rather than introducing an optimized quantum circuit, we build on related work that implements fermion hopping terms in the JW formalism, which will allow us to transfer some useful results on the parameter efficiency of our ansatz. We later define an ansatz without three-qubit gates by considering the rotation operators in the time-evolution directly.
We define rotation operators similar to the building blocks of the  parameter-efficient circuits of Ref.~\cite{gard2020efficient, ganzhorn2019gate, o2016scalable}, which maintain the particle number while defining rotations on two neighboring sites. The circuit implementing the  two-qubit entangling gate $A(\theta,\phi)$ is depicted in Fig.~\ref{fig:plaqcircuit_variational}a, when the two red $CCZ$ gates are discarded.
In the physical system, these two-qubit rotation gates operate within the single-excitation (one-electron) subspace, which is given by states of the form $\alpha \ket{01} + \beta \ket{10}$. In matrix form, this circuit performs a rotation given by
\begin{align}
    A(\theta, \phi) = \begin{pmatrix}
    1 & 0 & 0 & 0 \\
    0 & \cos\theta & e^{i\phi} \sin\theta & 0 \\
    0 & e^{-i\phi} \sin\theta & -\cos\theta & 0 \\
    0 & 0 & 0 & 1
    \end{pmatrix}\label{eq:Amat}
\end{align}
The circuit implementing $A(\theta, \phi)$ introduces three CNOT gates and two rotation gates, given by $R(\theta, \phi)$ (see Fig.~\ref{fig:plaqcircuit_variational}).

Just as in the case of the JWT, the bosonization procedure relates fixed-particle-number subspaces to fixed-excitation subspaces in the qubit Hilbert space. However, in order to work within the vacuum and periodicity constraints, we must adjust these rotations to correctly handle the auxiliary qubits as well. Therefore, we turn to the bosonized form of hopping operators. Following the procedure in Ref.~\cite{li2021higher}, we obtain the following form for the horizontal $T_{\vec{r}}^{\vec{x}} = f_{\vec{r}}^\dagger f_{\vec{r}+\vec{x}} + f_{\vec{r}+\vec{x}}^\dagger f_{\vec{r}}$ and vertical hopping $T_{\vec{r}}^{\vec{y}} = f_{\vec{r}}^\dagger f_{\vec{r}+\vec{y}} + f_{\vec{r}+\vec{y}}^\dagger f_{\vec{r}}$ 
\begin{align}
    T_{\vec{r}}^{\vec{x}} &\to  \frac{1}{2} \left(X_{\vec{r}}^{(1)} X_{\vec{r}+\vec{x}}^{(1)} + Y_{\vec{r}}^{(1)} Y_{\vec{r}+\vec{x}}^{(1)}\right) Z^{(2)}_{\vec{r}+\vec{x}}  \\
    T_{\vec{r}}^{\vec{y}} &\to  
    \frac{1}{2}\left(-X_{\vec{r}}^{(1)} Y_{\vec{r}+\vec{y}}^{(1)} + Y_{\vec{r}}^{(1)} X_{\vec{r}+\vec{y}}^{(1)}\right) Y^{(2)}_{\vec{r}}X^{(2)}_{\vec{r}+\vec{y}} 
\end{align}
Hence, in addition to the bosonic hopping operations (between brackets), we need to embed additional operations on the auxiliary system of the corresponding link, strictly \emph{within} the single-excitation subspace of the physical system. Circuits that meet this criterion are given in Fig.~\ref{fig:plaqcircuit_variational}. We explain the components of the vertical hopping along the $\vec{y}$ axis in more detail. We start from the operator $A(\theta, \phi)$ in Eq.~\eqref{eq:Amat}. In order to apply $Y^{(2)}_{\vec{r}}X^{(2)}_{\vec{r}+\vec{y}} $ in the single-excitation subspace, we insert double conditional $X$ and $Y$ gates that target the auxiliary qubits. The sign, on the other hand, is obtained by inserting an additional CNOT and conditional $Y$ gate in the physical system.  We denote the circuit operators in Fig.~\ref{fig:plaqcircuit_variational}a and \ref{fig:plaqcircuit_variational}b as $V_{\vec{x}}$ and $V_{\vec{y}}$ respectively. We define one layer $V_A$ of our variational ansatz as a set of $V_{\vec{x},\vec{y}}(\theta,\phi)$ circuits with varying angles $(\theta, \phi)$ applied to all pairs of qubits connected by an edge. This can be done with a similar layered circuit as in Ref.~\cite{gard2020efficient}, where it has been shown that such an approach spans the whole Hilbert space parameter efficiently.

\section{VQE}\label{sec:vqe}
\begin{figure}[tb]
\centering
\includegraphics[width=0.45\textwidth]{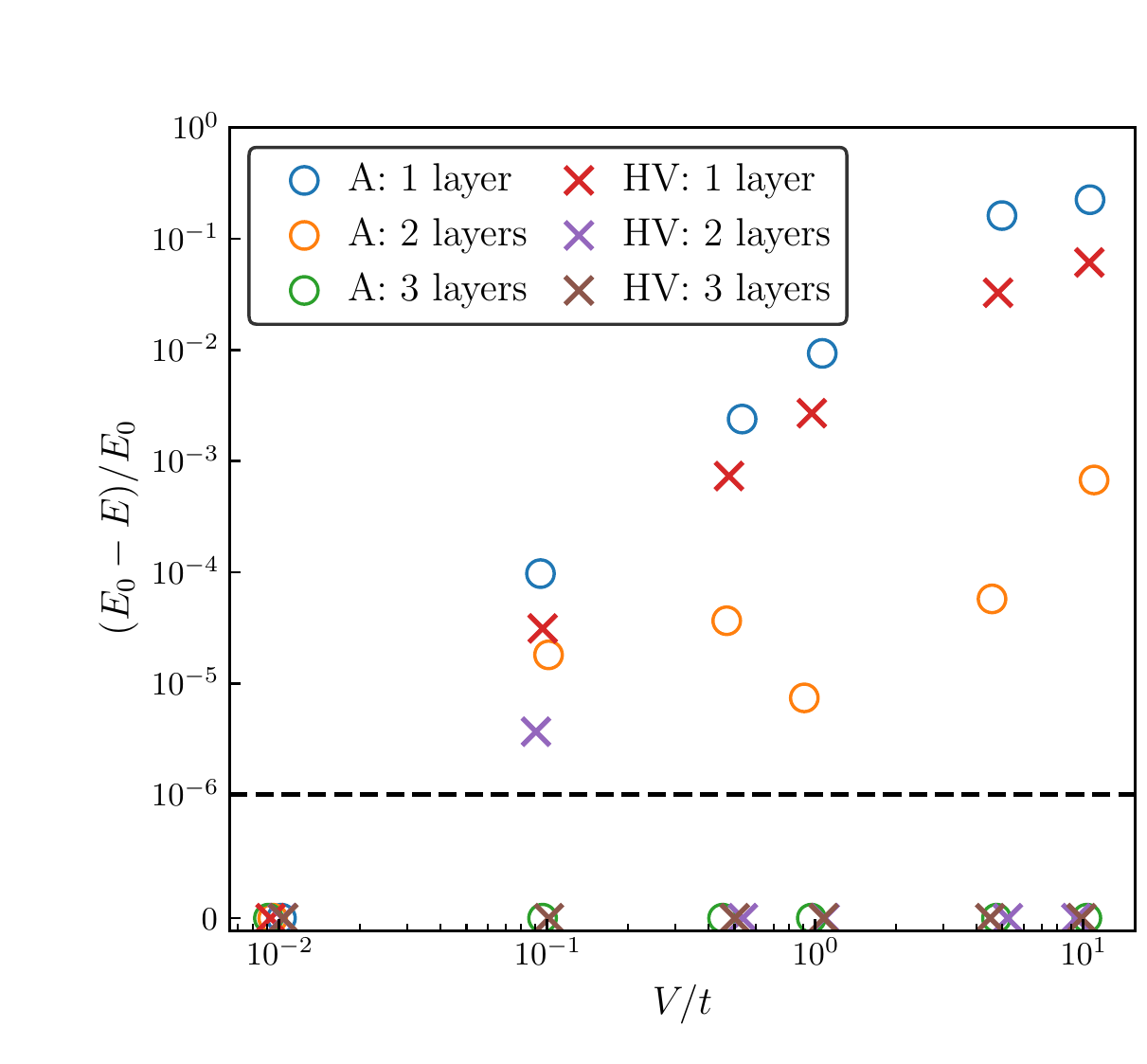}
\caption{Relative error on the ground state energy approximation using the method based on $A$ gates in Section~\ref{sec:variational_ansatz}, and using the HV ansatz with the circuit described in Section~\ref{sec:time_evo}. The HV ansatz operates on the free-fermion state ($V/t = 0$). The results are shown for a varying  number of layers and are compared to the exact ground-state energies. A small amount of jitter has been added to the $V/t$ values to show overlapping points. Relative energies below $10^{-6}$ have been set to $0$. Results are not shown for $V/t = 0$, since they are exact (up to rounding errors). Exact energies are obtained with NetKet~\cite{vicentini2021netket} \label{fig:2by4_result}}
\end{figure}

We test our variational circuit as a state ansatz in a Variational Quantum Eigensolver (VQE). Hereby, we insert two fermions and distribute them over the lattice by applying sequentially vertical parametrized hopping operators $V_{\vec{y}}$ on all vertical edges and horizontal hopping operators $V_{\vec{x}}$ on all horizontal edges. We take a separate set of gate parameters $(\theta, \phi)$ for each operator $V_{\vec{x},\vec{y}}(\theta, \phi)$.
In our simulations, we consider different values for the repulsion strength $V$ in Eq.~\eqref{eq:hamiltonian_fermion}, and consider up to three layers of the circuit. The results are shown in Fig.~\ref{fig:2by4_result}, where we compare the obtained energies with the exact ones for the $t$-$V$ model on a $2\times 4$ square lattice with a filling of $0.25$. Using two layers of the $A$-gate-based circuit, we obtain accurate approximations of the ground-state energies over the whole range of interaction strengths. Three layers provide essentially exact energies. The same results are obtained for a $3\times3$ lattice in Appendix~\ref{sec:oddbyodd}.
The time-evolution circuit defined in Section~\ref{sec:time_evo} can also be used to implement the HV ansatz if an initial state is given. We also present the results of this analysis in Fig.~\ref{fig:2by4_result} for a range of circuit depths, where the initial state is the free-fermion eigenstate. Two layers of the HV circuit are sufficient to obtain high-accuracy estimates of the ground states across the $V/t$ range, and three layers are essentially always exact.
These results are obtained by using the Adam optimizer, a learning rate of $10^{-4}-10^{-5}$, and by considering ideal, noiseless simulations. We do not find the need to use multiple parameter initializations and find good energies with each run.

\section{Scaling and comparison to other methods}\label{sec:scaling}
We analyze and compare the implementation cost of our method based on the locality of operators and the scaling behavior of the circuit depth and gate count with respect to the system size. For simplicity, we restrict to 2D square lattices and take $L=L_x=L_y$.
We separate two distinct parts in our circuit: the (1) initial-state preparation, which includes the vacuum state preparation and creating fermion pairs (see Section~\ref{sec:vacuum_prep}), and (2) the variational circuit (see Sections~\ref{sec:variational_ansatz}). Thirdly, we will discuss the cost of carrying out time evolution.

A thorough analysis of strategies for solving the fermionic systems on a lattice on near-term quantum computers was recently given in Ref.~\cite{cade2020strategies}. In their analysis, Cade~\etal compared the number of required gates of various methods by neglecting the cost of one-qubit gates and assuming that one can implement arbitrary two-qubit rotations. We will recapitulate the leading order behavior here.
We consider a horizontally snaked JW mapping. Hereby, the most challenging terms to implement are the vertical hopping terms $\exp[-i \theta (f_{\vec{r}}^\dagger f_{\vec{r}+\vec{y}} + f_{\vec{r}+\vec{y}}^\dagger f_{\vec{r}})] \to \exp[-i \theta (X_{\vec{r}} X_{\vec{r}+\vec{y}} + Y_{\vec{r}+\vec{y}} Y_{\vec{r}}) (\otimes_{\vec{r}' \in P_{\vec{r}\leftrightarrow{\vec{r}+\vec{y}}}} Z_{\vec{r}'})]$. The problem arises due to the presence of JW strings $\bigotimes_{\vec{r}' \in P_{\vec{r}\leftrightarrow{\vec{r}+\vec{y}}}} Z_{\vec{r}'}$ of $\order{L_x}$ that includes all qubits on the path $P_{\vec{r}\leftrightarrow{\vec{r}+\vec{y}}}$ between qubits $\vec{r}$ and $\vec{r}+\vec{y}$ along the JW snake ordering. Kivlichan~\etal~\cite{kivlichan2018quantum} introduced a circuit of fermionic SWAP gates (FSWAP) to handle these non-local terms. Their idea is based on the fact that vertical hopping terms between certain sites at the edges of the grid can be implemented locally, and therefore, one can permute the qubit columns until each vertical hopping term consists of Jordan-Wigner-adjacent positions. Permuting the qubits requires $L_x$ operations of the permutation operators, which themselves consist of $\order{L_x L_y}$ operations. Hence, this leads to a set of depth $\order{L}$ and $\order{L^3}$ FSWAP gates. The gate count can be reduced to $\order{L^2}$ using the parity-basis transformation in Ref.~\cite{jiang2018quantum}.

Our circuit in Section~\ref{sec:vacuum_prep} for the preparation of the initial state was constructed such that it exactly fulfills the periodicity and Gauss constraints, which allows us to store and adjust parities locally during the remainder of the circuit. The vacuum-state preparation does not contain variational parameters and only involves operations on the auxiliary system. The main bottleneck currently stems from the $V_{G}$ operator. The latter effectively rotates the system into all the correct $G_{\vec{r}}$ eigenstates. This requires applying $V_{G_{\vec{r}}}$ sequentially to all plaquettes, leading to a (sparse) $\order{L^2}$ total depth, and a depth of $\order{1}$ per register.
The number of gates applied therefore also scales according to $\order{L^2}$.  Similarly, Ref.~\cite{jiang2018quantum} requires the sequential application of CNOT parity operations to map a Jordan-Wigner-transformed system to the parity basis (yet, these can be implemented column-wise). For the total initial-state preparation, which includes both the vacuum-state preparation, fermion creation, and a layer of the variational circuit in Fig.~\ref{fig:plaqcircuit_variational}a and \ref{fig:plaqcircuit_variational}b, this means that the resulting circuit is $\order{L^2}$ depth and gate count. 

The initial state in the HV ansatz is required to be the ground state of a (typically non-interacting) Hamiltonian. The scaling of the circuit depth of our method is on par with the methods studied in Cade~\etal~\cite{cade2020strategies}, who found that the most efficient method of direct preparation of the initial state using Givens rotations~\cite{jiang2018quantum}, which has circuit depth $\order{L^2}$. 
However, Ref.~\cite{jiang2018quantum} also studies a fermionic Fourier transformation algorithm that becomes more depth-efficient for lattices larger than $20\times 20$, with circuits of $\order{L}$ depth, yet with $\order{L^3}$ gates~\cite{cade2020strategies}. This is the minimum depth required for
quantum information to be fully distributed across the qubit grid.

\begin{figure}[tb]
    \centering
    \includegraphics[width=0.40\textwidth]{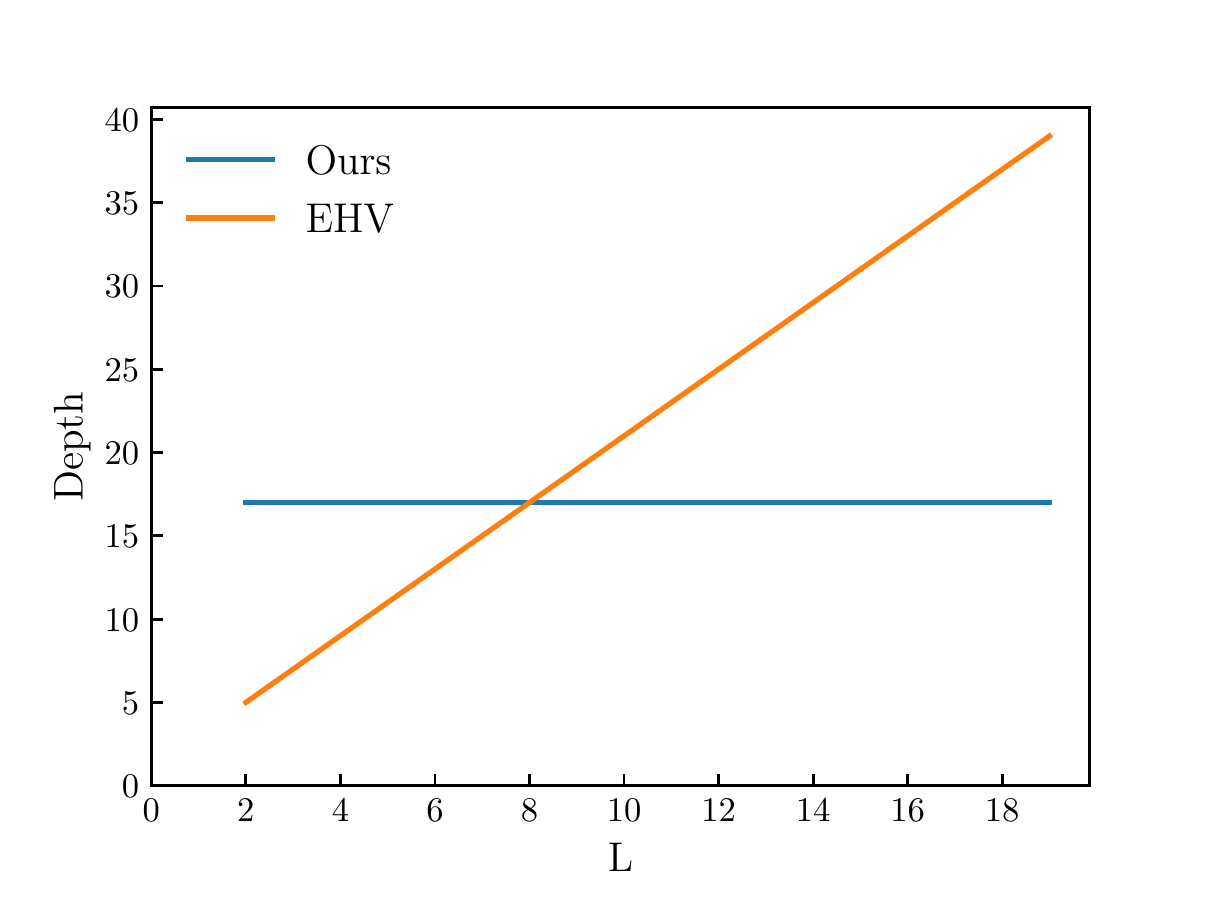}
    \caption{Comparison of the approximate depth of a circuit simulating a single Trotter step in the time-evolution of a fermionic system on a $L\times L$ square lattice, as a function of the system dimension $L$. We present a comparison between the efficient Hamiltonian variational ansatz (EHV)~\cite{cade2020strategies}, versus our local method.}
    \label{fig:depth_trotter}
\end{figure}

The variational part of our circuit includes only \emph{local} operations. Hereby, for both the horizontal and vertical hopping terms,  operations can be carried out in parallel~\footnote{When we say one can apply operations in parallel, this means one might need a constant depth to apply all operations. For example, to implement horizontal hopping terms, one can apply first all hopping terms to the even-numbered columns with the right to their right, and apply the same operations to the odd-numbered columns in the next layer. Since this has been illustrated in various other works (e.g.~Ref.~\cite{cade2020strategies}), we do not repeat the method here.}, leading to a constant depth and a gate count of $\order{L^2}$. Similarly, one step in the Trotterized time evolution requires constant depth with a gate count of $\order{L^2}$ local gates. Furthermore, compared to other work, we do not require networks of FSWAP gates to compute the correct parity factors in the time evolution.

The efficient Hamiltonian variational (EHV) ansatz requires a depth of $\order{L}$ and $\order{L^3}$ gates for a single Trotter step (mainly FSWAP gates). In Fig.~\ref{fig:depth_trotter}, we compare the depth of the EHV versus our method, where an arbitrary two-qubit rotation gate counts as depth 1, and a single-qubit gate does not contribute to the cost (as in Ref.~\cite{cade2020strategies}). We observe that a Trotter step becomes more efficient to simulate with our approach from system sizes greater than $8\times 8$.
The gate count of a Trotter step in the JW formalism can be further reduced to $\order{L^2}$ gates using the parity transformation in Ref.~\cite{jiang2018quantum}. Hence, while the gate count can be reduced to the same leading order in $L$, the depth of our circuit is $\order{L}$ more efficient.
In bosonization procedures and auxiliary fermion mappings the hopping terms are local in their qubit representation by construction. 
Since we store the parity locally in the vacuum state, we do not require any parity-related transformations based on FSWAP networks or sequential CNOT gates during the time-evolution (in contrast to other methods) and thereby reduce the depth of the Trotter step. 
Furthermore, generalizing the approaches based on FSWAP gates or CNOT parity transformations to efficient circuits for non-square lattices is non-trivial. Our approach can be generalized in a straightforward manner, since it only requires  following the bosonization recipe in Refs.~\cite{po2021symmetric} and defining the required local gauge transformations (see also Section~\ref{sec:generalization}). The time-evolution circuit can be further compressed using similar techniques as in Ref.~\cite{clinton2021hamiltonian}.

Finally, the last part is the implementation of measurements of Hamiltonians to determine the energy of the generated (variational) state. A naive direct implementation of the Hamiltonian would require measuring $\order{L^2}$ hopping and interaction terms, which in the JW formalism consist of non-local long-range JW strings involving $\order{L}$ qubits. This would render measurements practically challenging for higher-dimensional systems and increasing system sizes. Such a measurement implementation of our method would similarly include measuring $\order{L^2}$ terms, where each term is, however, now local by construction and involves only a constant number of qubits. The latter makes the naive approach feasible even for large systems and higher dimensions. Recently, various works have focused on devising techniques for more efficient measurements, which allow one to reduce the number of terms to measure independently, based on commuting sets~\cite{huggins2021efficient, crawford2021efficient, gokhale2020n, izmaylov2019revising, gokhale2019minimizing}. However, these methods do not solve the challenging problem of applying gates to non-JW-adjacent qubits. Furthermore, such techniques can also be directly applied to our method to reduce the number of independent measurements. Another method that would avoid measurements involving Jordan-Wigner strings is provided in Ref.~\cite{cai2020resource}. In this work, Cai describes a method based on SWAP gates similar to the approach to implement the vertical hopping terms using a similar reordering technique as described above until the measured operators are JW local. As we demonstrated, this adds a cost of $\order{L^3}$ gates and depth $\order{L}$.

\section{Generalization}\label{sec:generalization}
The Gauss-law constraints are referred to as `kinematic', meaning that they only depend on the lattice topology and the number of internal spin degrees of freedom. More specifically, the constraints depend on the edge attachment number, referring to the number of edges attached to each site ($4$ in the 2D square lattice studies in this work).
Hence, interestingly, for different Hamiltonians of systems restricted to the same lattice topology, the Gauss constraint remains the same, and hence, the same quantum circuit as in Figs.~\ref{fig:plaqcircuit_initial} and \ref{fig:plaqcircuit_variational} can be used in simulations.
As the dimensionality and structure of the lattice changes, the $G_{\vec{r}}$ operators become more complex (yet they remain spatially local). We leave the adaption of our method to these more involved cases for future work.

As the number of internal degrees of freedom increases, the number of physical qubits per site also increases. If we keep the same lattice structure (and hence keep the auxiliary fermion system the same), the  operator $C_{\vec{r}}$ in Eq.~\eqref{eq:gaussCdef} remains of Wen's plaquette model type, but the dynamical part of Eq.~\eqref{eq:gaussrepr} (r.h.s.) now contains $Z$ operators on the additional set of physical qubits (see Eq.~(21) in Ref.~\cite{li2021higher}).

As mentioned in Section~\ref{sec:introduction}, the local constraint in Eq.~\eqref{eq:gausslaw} can be interpreted as a Gauss law of a Chern-Simons lattice gauge theory~\cite{chen2018exact}. Our methodology can therefore be generalized to allow one to simulate such lattice gauge theories on a quantum device, by exploiting our approach to exactly satisfy the Gauss constraint. For other lattice gauge theories, similar methods as the one in this work have been explored, where the vacuum satisfies the gauge invariance condition~\cite{kaplan2020gauss, mazzola2021gauge}. 

\section{Conclusions and outlook}
Mapping fermion operators onto qubit degrees of freedom is not unique. In particular, transformations exist that map local fermion operators onto  local qubit operators. These mappings maintain locality by enlarging the Hilbert space using auxiliary degrees of freedom. However, in order to maintain equivalence between both representations, additional constraints must be satisfied, which again reduces the Hilbert space.
We introduce a strategy to design quantum circuits that exactly fulfill these constraints. We provided a circuit that implements the time-evolution operator of a local Hamiltonian with purely local gates and found that it requires only $\order{1}$ circuit depth per Trotter step, which is an improvement compared to other work that requires $\geq \order{L}$ depth.
We finally study variational quantum circuits that maintain the fermion-number and periodicity constraints. We demonstrate how ground-state energies of a local fermion Hamiltonian, i.e.\ the $t$-$V$ model, can be obtained in combination with VQE.

In future work, we will explore implementations of fermionic Fourier transformations for initial-state preparation. Furthermore, using different vacuum-state preparation techniques, it might be feasible to reduce the circuit depth of the vacuum-state preparation circuit to $\order{L}$ using column-wise parity rotations as in Ref.~\cite{jiang2018quantum}. Also, while the focus of this work was to introduce our method, one can search for lower-cost and more hardware-efficient circuits than the ones presented here. Another clear extension is to consider different lattice topologies, where in the case of triangular lattices, one can build on work that studies circuits to prepare Kitaev honeycomb ground states (such as in Ref.~\cite{bespalova2021quantum}). We also foresee extensions that aim at reducing the number of required auxiliary qubits. Finally, our method can be extended to the wide range of local fermion-to-qubit mappings currently available~\cite{setia2018bravyi, setia2019superfast, chen2018exact, chen2022equivalence, bochniak2020bosonization, po2021symmetric, derby2021compact}.

\section*{Acknowledgement}
This work was supported by Microsoft Research.

\section*{Code availability}
A code implementing all the quantum circuits discussed in this work has been made available at \url{https://github.com/cqsl/local-f2q-circuit} (in Pennylane~\cite{bergholm2018pennylane}).

\bibliographystyle{quantum}
\bibliography{biblio}% Produces the bibliography via BibTeX.

\appendix

\section{Additional experiments}\label{sec:oddbyodd}
To demonstrate the technique on an odd by odd lattice, we run the experiments in Fig.~\ref{fig:2by4_result} for $L_x = L_y = 3$, which requires simulating $18$ qubits (of which $9$ physical and $9$ auxiliary). In the main text we have assumed even-by-even lattices for simplicity. In general, an additional on-site parity factor $\rho$ must be taken into account, which has tacitly been set to $+1$ in the main text (see Ref.~\cite{po2021symmetric} for a detailed discussion). The horizontal hopping term reads
\begin{align}
    f^\dagger_{\vec{r}} f_{\vec{r}+\vec{x}} + f^\dagger_{\vec{r}+\vec{x}} \to& \frac{\rho}{2}   ( X^{(1)}_{\vec{r}}X^{(1)}_{\vec{r}+\vec{x}} + Y^{(1)}_{\vec{r}}Y^{(1)}_{\vec{r}+\vec{x}})Z^{(2)}_{\vec{r}+\vec{x}} \label{eq:spinless_ham_boson_rho}
\end{align}
and the global periodicity constraints become,
\begin{align}
     \rho^{L_x}\oprod_{m=0}^{L_x-1} Z^{(1)}_{\vec{r}+m\vec{x}} Z^{(2)}_{\vec{r}+m\vec{x}} &\overset{c}{=} -1 \label{eq:pbc_constraint_x_rho} \\      
     (-1)^{L_y}\oprod_{m=0}^{L_y-1} Z^{(2)}_{\vec{r}+m\vec{y}} &\overset{c}{=} -1 \label{eq:pbc_constraint_y_rho}
\end{align}
while the local constraint in Eq.~\eqref{eq:gaussrepr} remains the same.

\begin{figure}[htb]
\centering
\includegraphics[width=0.45\textwidth]{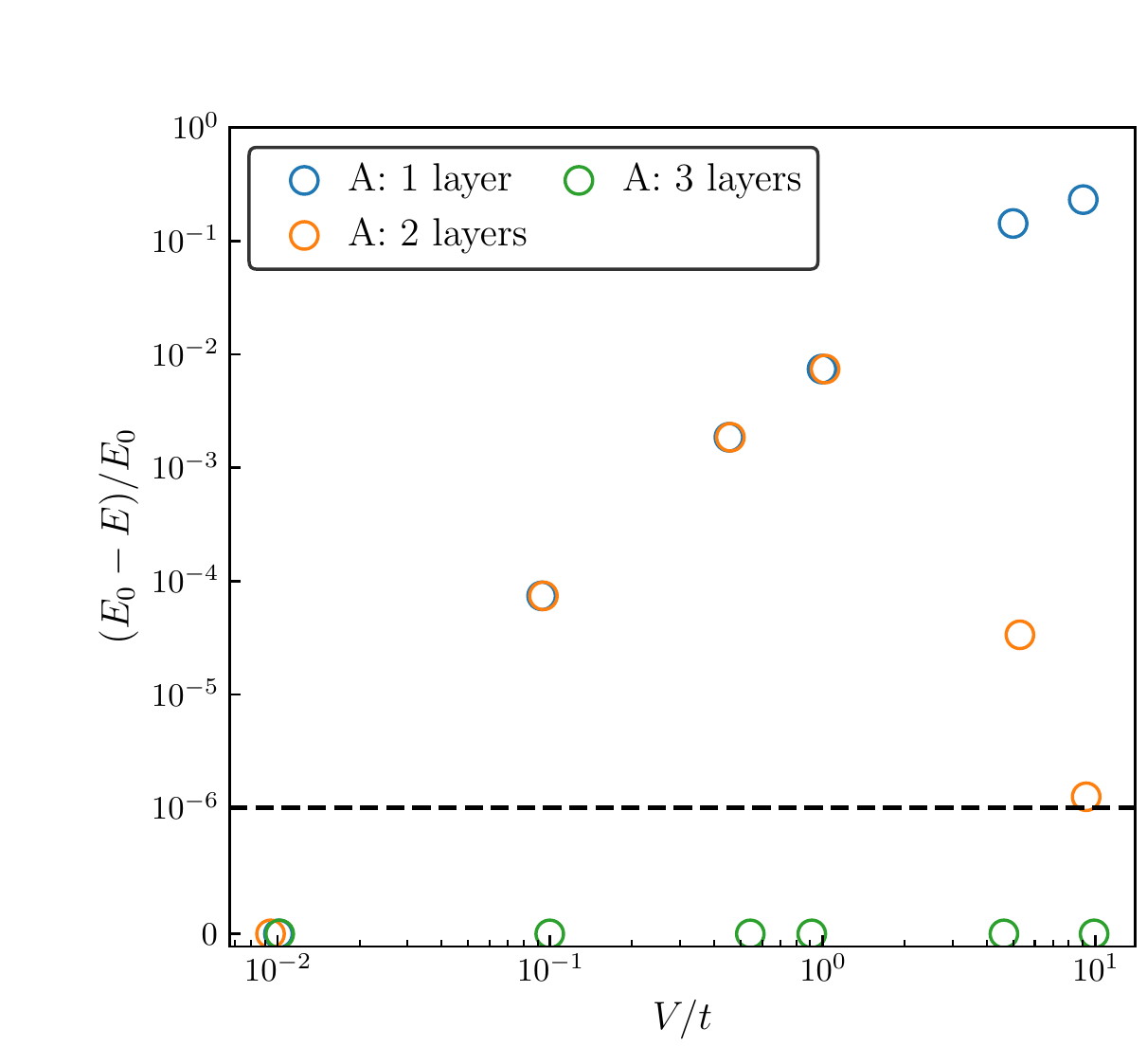}
\caption{Same as in Fig.~\ref{fig:2by4_result}, but for a $3\times3$ lattice with APBC.\label{fig:3by3_result}}
\end{figure}

These constraints can be satisfied by setting $\rho = -1$ and changing the transformation in Fig.~\ref{fig:periodicity_circuit} to the identity operator (i.e.\ not applying any $X$ gates). The results for the $3\times3$ lattice are shown in Fig.~\ref{fig:3by3_result} and lead to the same conclusions as in the main text, showing that two layers of the variational ansatz already lead to good representational power across the range of couplings.

% \newpage
	% \bibliography{biblio.bib}
\end{document}